\documentclass[conference]{IEEEtran}
\IEEEoverridecommandlockouts
\usepackage{bm}
\usepackage{amsmath,amssymb,amsfonts}
\usepackage{pifont}
\usepackage{algcompatible}
\usepackage[ruled]{algorithm2e}
\usepackage{algpseudocode}
\usepackage{steinmetz}
\usepackage{graphicx}
\graphicspath{{sims/}}	
\usepackage{textcomp}
\usepackage{color,xcolor}
\usepackage{lipsum}
\usepackage{array}
\usepackage{multirow}
\usepackage{url}

\usepackage[numbers,sort&compress]{natbib}

\newcommand{\ignore}[1]{}

\newcommand{\NFMA}[0]{{N_{\mathrm{FMA}}}}
\newcommand{\pout}[0]{{p_{\mathrm{out}}}}
\newcommand{\neab}[0]{{n_{\mathrm{eab}}}}
\newcommand{\necb}[0]{{n_{\mathrm{ecb}}}}

\newcommand{\pin}[0]{{p_{\mathrm{in}}}}

\newcommand{\ceil}[1]{\left\lceil #1 \right\rceil}
\newcommand{\floor}[1]{\lfloor #1 \rfloor}

\DeclareSymbolFont{symbolsC}{U}{ntxsyc}{m}{n}
\SetSymbolFont{symbolsC}{bold}{U}{ntxsyc}{b}{n}
\DeclareMathSymbol{\multimapdotbothB}{\mathrel}{symbolsC}{24}

\newif\ifShowCorrections
\ShowCorrectionstrue
\ifShowCorrections
\usepackage[normalem]{ulem}
\definecolor{forgreen}{rgb}{0,0.6,0}
\definecolor{orange}{RGB}{255,140,0}
\definecolor{skyblue}{RGB}{100, 150, 235}

\newcommand{\mmc}[1]{{\color{forgreen}[#1]}}

\newcommand{\swc}[1]{{\color{orange}[#1]}}
\else

\newcommand{\mmc}[1]{}

\newcommand{\swc}[1]{}
\fi
\algnewcommand\algorithmicfore{\textbf{for}}
\algdef{S}[FOR]{For}[1]{\algorithmicfore\ #1\ \algorithmicdo}

\begin{document}

\title{Generalized Methodology for Determining Numerical Features of Hardware Floating-Point Matrix Multipliers: Part I
}

\author{\IEEEauthorblockN{Faizan A.~Khattak and Mantas Mikaitis}
  \IEEEauthorblockA{School of Computer Science, University of Leeds, Leeds, UK
  }}

\maketitle

%
%
\begin{abstract}
  Numerical features of matrix multiplier hardware units in NVIDIA and AMD data centre GPUs have recently been studied. Features such as rounding, normalisation, and internal precision of the accumulators are of interest.
  In this paper, we extend the methodology for analysing those features, to consumer-grade NVIDIA GPUs by implementing an architecture-independent test scheme for various input and output precision formats. Unlike current approaches, the proposed test vector generation method neither performs an exhaustive search nor relies on hard-coded {constants that are device-specific, yet remains applicable to a wide range of mixed-precision formats.
We have applied the scheme to the RTX-3060 (Ampere architecture), and Ada RTX-1000 (Ada Lovelace architecture) graphics cards and determined numerical features of matrix multipliers for binary16, TensorFloat32, and bfloat16 input floating point formats and binary16 and binary32 IEEE 754 output formats.
Our methodology allowed us to determine that} the numerical features of RTX-3060, a consumer-grade GPU, are identical to those of the A100, a data centre GPU.
We do not expect our code to require any changes for performing analysis of matrix multipliers on newer NVIDIA GPUs, Hopper or Blackwell, and their future successors, and any input/output format combination, including the latest 8-bit floating-point formats.
\end{abstract}

\section{Introduction}
As the demand for greater computational efficiency in machine learning continues to rise, low-precision arithmetic has emerged as a key method for training neural networks faster.
Recent advancements have pushed precision levels down to as low as $4$ bits on modern GPUs~\cite{nvid25b, amd25}.
Majority of modern GPUs are equipped with dedicated matrix multiplication units, designed to accelerate linear algebra operations---particularly dense matrix multiplication, which is central to both training and inference in deep learning workloads.
These units also support low-precision arithmetic, thereby enabling significant throughput improvements.
However, $+$ and $\times$ operations implemented within them typically do not conform to the IEEE 754 standard~\cite{ieee19}, and their numerical behaviour is seldom documented and differs between architectures.
IEEE 754 reduction operations are implementation-defined~\cite[Sec.~9.4]{ieee19}.
The goal of this work is to characterise the numerical features of these matrix multipliers.
The characterisation not only supports the standardisation efforts of the IEEE P3109 working group for floating-point arithmetic in machine learning~\cite{ieee25} but also benefits scientific computing community that traditionally relies on IEEE 754 binary64 hardware, by helping them interpret differences in results computed on various platforms.

Previously, {Hickmann~and~Bradford~}\cite{hibr19} have reported various numerical features extracted via hard-coded  test vectors, thereby estimating the features of NVIDIA V100 \emph{tensor cores}, a term used by NVIDIA for referring to matrix multipliers.
Building upon their work, Fasi,~Higham,~Mikaitis,~and~Pranesh~\cite{fasi2021} have {analysed} additional features, such as {the number of} extra bits for carries {in the intermediate accumulator} and monotonicity of the dot product, by adopting a similar strategy of generating test vectors via constant parameters selected for each case of input and output precision format.
They have reported features of matrix multipliers of the V100, T4, and A100 GPUs based on the Volta, Turing and Ampere architectures of NVIDIA for a variety of floating point formats, such as binary16,~bfloat16, TensorFloat32, binary32, and binary64.
Li et al.~\cite{llfs24} have subsequently applied the techniques on the AMD and NVIDIA H100 GPUs, and proposed a partially generalized test vector generation scheme which can be applied to various input and output floating-point precision formats.
However, due to limited number of features considered, the proposed algorithms, specifically for determining fused-multiply-accumulate (FMA) width therein, seem to be inapplicable to all GPU architectures~\cite{llfs24}.
Finally, the satisfiability modulo theories (SMT) based work conducted by Valpey~et al.~\cite{vlpg25} performs an exhaustive or close to it search through the input space to determine features of matrix multipliers of NVIDIA GPUs.
We demonstrate algorithms that can determine many numerical features efficiently without needing to traverse the large input space of floating-point numbers that methods based on SMT need to do; for instance Valpey et al.~\cite{vlpg25} reported that it took six hours to determine a test vector for the number of extra carry bits in the Ampere architecture but still failed to converge.

The software released circa 1982, \emph{Paranoia}\footnote{\url{https://www.arithmazium.org/paranoia/aaapara_toc.html}}, made for testing machines' arithmetic behaviour before IEEE 754 standardisation, inspired this work.

Our contributions are as follows.
\begin{enumerate}
\item Previous studies have focused on data centre GPUs whereas this work targets consumer-grade GPUs,
\item the proposed approach supports varying input and output precision parameters, making it readily applicable to many available input-output format combinations,
\item this work highlights internal dependencies among numerical feature tests and presents an example demonstrating how one numerical feature test can influence the tests for others,
\item the proposed test vector generation model is architecture-agnostic and thus applicable across different GPU architectures—unlike the approach of Li et al.~\cite{llfs24}, which, despite partial generalization, is applicable to Ampere but may not apply to Volta architecture without manual code changes, and
\item we determine the relationship between the numerical features of consumer-grade GPUs and those of data centre GPUs based on the same architectures.
\end{enumerate}

\section{Model, Parameters and Numerical Features}
\subsection{Model for Test Vector Generation}
Matrix multiplication on the latest NVIDIA GPUs performs matrix-multiply accumulate operation
\begin{align}
D=AB+C \in \mathbb{R}^{m\times n},
\end{align}
where $A \in \mathbb{R}^{m\times k}$, $B \in\mathbb{R}^{k\times n}$.
For large matrices, multiplication is accomplished by partitioning the given matrices into smaller-sized sub-matrices which are called \emph{tiles}.
Computations from tiles are aggregated to produce the final result.
Each tile may be computed by several tensor cores, which operate on even smaller blocks.
In order to examine the numerical behaviour of these matrix multipliers that work at the tile level, an analysis of a single element of $D$ is sufficient.
Therefore, an element of $D$ at $i$th row and $j$th column can be represented as
\begin{align}
\label{eq_d11}
d_{ij}=\sum_{\ell=1}^{k}a_{i\ell}b_{\ell j}+c_{ij}.
\end{align} 
We do not need to refer to separate elements in $D$, so we have
\begin{align}
\label{eq_d1}
d=\sum_{l=1}^{k}a_{l}b_{l}+c=\sum_{l=1}^{k}r_{l}+c,~\text{where}~r_{l}:= a_{l}b_{l}.
\end{align}

\subsection{Definitions}
We define {\emph{FMA width}} following Li~et~al.~\cite{llfs24}, as the minimum number of multiply-accumulate operations in~\eqref{eq_d1} before rounding and normalisation of the accumulator, to the output floating-point format, is applied, and we denote it with $\NFMA$.
This parallel fashion FMA is also termed as block FMA~\cite{bhlt20}.
Block FMAs may perform accumulation in higher precision than the output format.
Therefore these units increase not only performance but may also increase accuracy, because of the precision growth due to carry bits and single normalisation at the end of sum of length $k$~\cite{mika24}.
In a block FMA, the final conversion to the output format may be deferred until all $k$ products are added.
The number of input and output format significand bits are denoted by $\pin~\text{and}~\pout$, respectively~\cite{ieee19}.
It is reminded that the scope of definition of $\pout$ is limited to binary32 and binary64 floating-point formats because element-wise multiplication for binary16, TensorFloat32, and bfloat16 block FMAs in Volta and Ampere takes place at least in binary32 precision, and thereafter rounding is applied to cast to lower precision if required~\cite{fasi2021}.
Therefore we have $\pout > 2\pin$ {but the tests given below are {applicable}
  to $\pout>(\pin+n_{\mathrm{ecb.max}})$
  where $n_{\mathrm{ecb.max}}$ is defined as the maximum number of extra carry bits (see~\eqref{eq_ub})}.
We define \emph{extra carry bits} as additional bits that support carry propagation across successive additions in \eqref{eq_d1} without requiring immediate normalisation after each step. For instance, consider the addition $1.01 + 1.00 = 10.01,$
where the result is left denormalized. This intermediate result is then added to another number:
$
10.01 + 1.01 = 11.10.
$
Only after the full accumulation is the result normalized to:
$
1.11 \; \text{(with the exponent appropriately adjusted)}.
$
This behavior---of deferring normalisation and preserving carry information without reducing precision---is what we refer to as having an \emph{extra carry bit}.
{
  If an implementation produces $1.01+1.00=10.01$ but normalizes and rounds the result to $1.00$ before passing it to the next addition then zero extra carry bits are available in the accumulator's precision.
\subsection{Scope of Numerical Features}
The scope of this paper encompasses a broad range of numerical features, many of which have been investigated in prior research, however only for data centre GPUs. 
First, the support for subnormal numbers in both input and output is examined.
Next, the size of the accumulator is determined by identifying the number of additional bits allocated for aligning significands, as well as those reserved to accommodate carries generated during the accumulation process.
In addition, the FMA width is determined for inputs in various floating-point precision formats using an iterative algorithm.
Moreover, the rounding mode is examined for outputs in {binary16}, {bfloat16}, {TensorFloat32}, and {binary32} precision formats, including scenarios where results from two different block FMA operations are accumulated.
Finally, the dependencies among numerical features are highlighted using an example algorithm that produces correct results only when the extra bits for carries and significand alignment are determined beforehand.   

\section{Generation of Test Vectors}
We formulate expressions for the generation of each test vector or a series of test vectors needed to reveal a particular numerical feature of matrix multiply hardware units.
The number of precision bits in the input and output formats is the input to expression forming rules.
This feature can help apply these tests in any input and output floating point formats that are available now or may become available in the future.

\subsection{Subnormal Support} 
Subnormal number support for both input and output can be verified using simple test cases, as shown by Fasi et al.~\cite{fasi2021}. Since prior works~\cite{llfs24, fasi2021, hibr19} agree on subnormal support in data centre GPUs, our motivation for revisiting this feature is to assess whether consumer-grade GPUs offer similar support. Accordingly, we reapply the test methods from~\cite{fasi2021, hibr19, llfs24} to selected consumer GPUs.

\subsection{Rounding Modes}

This section addresses the rounding mode of each addition operation in (3) as well as the final rounding to output precision.
The standard~\cite{ieee19} rounding modes are RoundToNearest (RN), RoundTowardsZero (RZ), RoundUp (RU), RoundDown (RD). 
Fasi et al.~\cite{fasi2021}, have generated test vectors that indicated RZ as the rounding mode in addition in several block MMA designs, and Valpey et al.~\cite{vlpg25} have built on their work to demonstrate that block FMAs do not provide results consistent with any of the standard rounding modes for subtraction.
In the former, the example test provided for the V100 GPU is $2+(\frac{3}{4}\times 2^{-22})$, which is reported to result in $2$ (and similarly, $-2$ on the negative axis), and is concluded to be consistent with RZ rounding mode for adding two positive numbers.
Valpey~et~al.~\cite{vlpg25} additionally used $2-2^{-41}$ to demonstrate that $-2^{41}$ is not preserved in the alignment step, consistent with the behaviour of RU for subtraction.
Combining Fasi et al.~\cite{fasi2021} result and their new test with subtraction they were able to conclude that block FMAs do not correspond to any standard rounding mode, consistent with bit truncation.

It is important to define \emph{truncation} precisely as one can truncate in the significand alignment step or after the addition result is obtained with or without extra bits.
We rely on the model assumed in~\cite{hibr19} where the alignment step is expected to be followed either by truncation or rounding before accumulation has begun. 
Once addition is performed within a block FMA, another truncation or rounding is performed to output the accumulated results in output precision.
To determine if there is any intermediate rounding post alignment, we need to know the number of extra alignment bits, which we denote by $\neab$.
With $\neab$ known and $\NFMA\ge 2$, we suggest $c=\pm 2^{j},~r_1=r_2=\pm(2^{-\pout+j-\neab}+2^{-\pout+j-\neab-1})$. 
The output $d$ must be $\pm 2^{j},\pm (2^j+2^{-\pout+j+2-\neab})$,~$\{2^j+2^{-\pout+j+2-\neab},-2^j\}$, and $\{2^j,-(2^j+2^{-\pout+j+2-\neab})\}$ if signficand bits beyond the output precision are truncated, rounded via RN with ties-to-even (RNE), RU, and RD, respectively.
This is applicable for $\neab=0$ and $1$.
For $\neab>1$, more sophisticated tests are needed.

To determine the final rounding mode applied to the output of a block FMA, i.e.,  $\mathrm{r}\{c + r_1 + \dots + r_{\NFMA}\}$
where $\mathrm{r}\{\cdot\}$ denotes the rounding operation, we assume $\NFMA \geq 3$. 
This assumption is required when $\neab = 0$. 
For the case $\neab = 1$, it is sufficient that $\NFMA \geq 2$, while for $\neab > 1$, the weaker condition $\NFMA \geq 1$ is acceptable.
For simplicity, we assume $\neab=0$ because we are not sure if all modern GPUs allocate extra alignment bits in their matrix multiplier units.
Then we propose $c=\pm(2^j+2^{-\pout+j+1}+2^{-\pout+j+2})$ and $r_1=r_2=r_3=\pm(2^j)$. 
These input vectors are such that they can prevent intermediate truncation or rounding as none of the bits gets beyond the output precision---instead extra carry bits are utilized.
Since normalisation is delayed until a complete block FMA operation has been performed, we must have $d=\pm(2^{j+2})$, $\pm(2^{j+2}+2^{-\pout+j+3})$, $\{2^{j+2}+2^{-\pout+j+3},-2^{j+2}\}$, and $\{2^{j+2},-(2^{j+2}+2^{-\pout+j+3})\}$ for truncation, RNE, RU, and RD rounding modes, respectively.

\subsection{Features of the Accumulator}
Since the scope of this paper does not include the order in which features should be determined, we therefore rely on certain assumptions from earlier work in this area. 
\subsubsection{Bits for Significand Alignment}
\label{sec:extra-align-bits}

To reveal the number of extra bits allocated in the alignment of significands in the accumulator, 
we propose setting $c=2^j,j\in\mathbb{N}_0$, and applying the following tests:\\
$\bullet~$choose $a_{i},b_{i}$ $i\in\{1,2\}$ such that $r_{i=1,2}=2^{-\pout+j}$. Assuming products are exact when they leave multiplication units,
  if the resulting $d = 2^j+2^{-\pout+1+j}$, this implies the presence of at least an extra bit in the alignment. This test assumes no normalisation after a single binary operation, i.e., addition involving two terms.\\
$\bullet$ with $r_1=2^{-\pout+j}$ and $r_{i\in\{2,3\}}=2^{-\pout+j-1}$, and under the assumption of alignment of significands w.r.t. the largest exponent, we must still have $d=2^j+2^{-\pout+j+1}$ if two extra bits are utilized in the significands' alignment.\\
$\bullet~$to detect $n_{\mathrm{eab}}$ extra bits in alignment, we suggest
 $r_{i\in\{1,...,n_{\mathrm{eab}}-1\}}=2^{-\pout-i+j+1}$ and $r_{i\in\{n_{\mathrm{eab}},n_{\mathrm{eab}}+1\}}=2^{-\pout+1-n_{\mathrm{eab}}+j}$. 
 If $n_{\mathrm{eab}}<\NFMA$ and we have $d=2^j+2^{-\pout+j+1}$, this indicates $n_{\mathrm{eab}}$ extra bits are present. 
 This test assumes rounding is performed after a complete block FMA operation.{ The final result remains invariant under any rounding mode post normalisation because all bits beyond the last bit in the output precision are zero.}

In these tests, we assume that when significands are aligned, the bits beyond the last extra bit{, the $(\pout+\neab)$th bit,} are truncated { without any types of rounding. 
Instead rounding is applied only once when all terms within a single block FMA have been accumulated}.
The parameters $r$ should be chosen such that the target product values used in the tests can be accurately represented in the output precision.
\subsubsection{Normalisation}
In compliance with IEEE 754, normalization is applied after each addition operation. 
However, in the matrix multiplier units, present in graphics card of various vendors, this is not the case for various input precision formats~\cite{fasi2021, llfs24, hibr19}.
Multiple extra bits are reported to exist~\cite{fasi2021} when significands are aligned in floating point arithmetic for addition or subtraction, and also to accommodate carries in such multi-term addition to prevent the need for immediate normalization. 
Therefore, there is mutual connection between the need for normalisation and the presence of these extra bits---otherwise internal accumulator would overflow, which would deem the floating-point computation incorrect.
Hence in determining whether immediate or late normalisation occurs, we perform a test that takes into account all possible cases of extra carry and alignment bits. 
Before we proceed, we define $eab$ and $ecb$ to denote the presence of extra alignment and extra carry bits, respectively where each can take on a value $0$ or $1$ irrespective of the number of bits present.
On the basis of $\{ecb,eab\}$, we can have $4$ possible cases:
\begin{itemize}
\item $\{0,*\}$ Irrespective of whether there are alignment bits, when there are no extra carry bits, immediate normalisation must take place after each addition.
\item $\{1,0\}$ The mere presence of extra bits present to accommodate carries implies that immediate normalisation after every binary operation may not take place. 
To test this, we set $c ={2-2^{-\pout+1}},~r_{i\in\{1,2,3\}}=2^{-\pout+1}$. 
In case of immediate normalization, we must have $d=2$ assuming addends are aligned w.r.t. the largest exponent.
\item $\{1,1\}$ With $c =1-2^{-\pout+t}$, $t\ge 3$, $~r_{i\in\{1,2\}}=2^{-\pout+t}+2^{-\pout}$, we must have $d=1+2^{-\pout+t}$
  if normalisation is immediate with RZ/RD/RNE/truncation as rounding modes and $d=1+2^{-\pout+t}+2^{-\pout+2}$ for the case of RU;
  otherwise $d=1+2^{-\pout+t}+2^{-\pout+1}$. 
  The $t\ge 3$ helps create a separation of at least one bit between LSB and the consecutive ones in $c$ to help generate carry in the MSB.
\end{itemize}

\subsubsection{Carry Bits}
The number of extra bits needed in the accumulator to support carries is dependent upon the FMA size and the inner product normalisation algorithm.
If each binary addition is followed by immediate normalisation, then no extra carry bits are needed—according to our definition of extra carry bits, which aligns with the definition used in~\cite{fasi2021}.
To determine the number of extra carry bits, we propose a test that is unaffected by the presence of extra alignment bits, as there is no explicit dependency.
While such a dependency could arise depending on how test vectors are constructed, our proposed method remains free from it as outlined in Algorithm~\ref{alg_ECB}.
Additionally, we address the dependency on FMA size---since it dictates how many extra carry bits are required---by iteratively increasing the shared dimension of both $A$ and $B$, i.e., $k$ from $2$ until it exceeds the FMA size.
Once $k$ exceeds $\NFMA$, the condition of the if statement becomes true because the absolute value of $d$ is no longer equal to the absolute of sum of $c$ and $r_{i\in\{1,\dots,k\}}$ due to independent normalisation and rounding in each block FMA operation. 
Hence, the proposed algorithm resolves the dependency of the FMA size by iteratively increasing the shared dimension of input matrices $k$.

For determining the number of extra bits allocated to propagate carries from one addition to another without intermediate normalisation, we require $1$s in the MSBs as well as in the LSBs to detect the carry bits, while simultaneously keeping track of whether the LSB has been utilized instead of being truncated.
During accumulation, if $n$ carries occur in the MSBs, the same number of carries must be generated in the region of LSBs.
This is because, after normalisation, the result will be right-shifted by the same number of bits used to support the carries. 
Therefore, carry detection is performed using the last bit in the updated LSB after the precision has been reduced, and it has to be $1$ from the carry in the LSBs before the final normalisation.
Although such an algorithm is feasible, its output depends on the rounding mode.
Therefore, the FMA detection iterative algorithm breaks the while loop once $k$ exceeds the FMA size.
In the meantime, the $\necb$ kept on detecting the number of carries with the help of second if statement by the relation
\begin{align}
\label{eq_necb_k}
n_{\mathrm{ecb}}=\left\lfloor\mathrm{log}_2(k(2-2^{-\pin+1}))\right\rfloor,
\end{align}
where the term $2-2^{-\pin+1}$ is the closest value below $2$ in the accumulator’s internal precision with input precision $\pin$, ensuring the MSBs contain the longest run of $1$s to maximize carries with the fewest added terms. 
We assume that each product term is exactly representable as the product of two operands in the input precision.

\ignore{Let us assume an example case where $\NFMA=8$, not known in advance, and the Algorithm~\ref{alg_ECB} begins with $k=2<\NFMA$. 
The condition in the if statement is not satisfied until $k$ exceeds $8$.
For instance, for $k=9$, $r_{9}$ gets passed into a separate block FMA operation, and we obtain
\begin{align*}
d=\mathrm{nr}\left\{\mathrm{nr}\left\{c+\sum_{i=1}^{8}r_i\right\}+\mathrm{nr}\{r_9\}\right\}
\end{align*}
instead of 
$\mathrm{nr}\{c+\sum_{i=1}^{9}r_i\} 
$ where $\mathrm{nr}$ stands for normalisation followed by rounding.
This leads to inequality condition in the if statement to get satisfied, and the Algorithm~\ref{alg_ECB} terminates and outputs {$n_{\mathrm{ecb}}=3$ for binary16 input where $\pin=11$}.
Note}

\ignore{\begin{algorithm}[t]
\caption{Iterative approach for determining the number of extra carry bits and $\NFMA$.}
\label{alg_ECB}
\textbf{Output:} $\NFMA,\necb$\; 
$k=2$\;
\While{\textbf{true}}
{$r_{i\in\{1,\dots,k\}}=0$\;
$c=(2-2^{-\pin+1})+\sum_{i=1}^{\ceil{\mathrm{log2}(k)}}2^{-\pout+i}$\;
$r_{i\in\{1,\dots,(k-1)\}}=2-2^{-\pin+1},~r_{k}=2^{-\pout+1}$\;
\textbf{call~matrix~multiplier}\;
\If{$d\neq (c+\sum_{i=1}^{k}r_{i})$}
{
$\NFMA\gets (k-1)$\;
$n_{\mathrm{ecb}}=\mathrm{log}_2\left(\left\lceil \frac{2(k-1)}{2 - 2^{-\pin+1}} \right\rceil - 1\right)$\;
\textbf{return $\{n_{\mathrm{ecb}}, \NFMA \}$}\;

}

$k\gets (k+1)$
}

\end{algorithm}
}
\begin{algorithm}[t]
\caption{Iterative approach for determining the number of extra carry bits and $\NFMA$.}
\label{alg_ECB}
\textbf{Output:} $\NFMA,\necb$\; 
$\necb=0,~k=2$\;
\While{\textbf{true}}
{$r_{i\in\{1,\dots,k\}}=0,~r_{1}=\pm 1,~r_{k}=\pm 2^{-\pout+1}$\;
$c=\pm(1+2^{-\pout+1})$\;
\textbf{call~matrix~multiplier}\;
\If{$|d|\neq |c+\sum_{i=1}^{k}r_i|$}
{
$\NFMA\gets (k-1),~\textbf{break}$\;
}
$r_{i\in\{1,\dots,k-1\}}=2-2^{-\pin+1},~r_{k}=2^{-\pout+1}$\;
$c=2-2^{-\pin+1}+\sum_{i=1}^{\ceil{\mathrm{log_2}(k)}}2^{-\pout+i},~$\;
\textbf{call~matrix~multiplier}\;
\If{$d = (c+\sum_{i=1}^{k}r_i)$}
{
$n_{\mathrm{ecb}}=\floor{\mathrm{log_2}(k(2-2^{-\pin+1}))}$\;
}

$k\gets (k+1)$
}

\end{algorithm}


\subsection{FMA Size}
In determining the FMA size of a GPU matrix multiplier unit (or a tensor core in NVIDIA GPUs), we have to rely on the assumption that precision is preserved within a single block FMA operation, with rounding and/or normalisation deferred until the final result~\cite{llfs24, bhlt20}. 
The iterative algorithm presented by Li et al.~\cite{llfs24} appears to have a typographical error, as its behaviour does not align with the reference implementation provided on the FPTalk24 website~\cite{li24}. 
The implementation employs two non-zero products, whereas the pseudo-code in~\cite{llfs24} relies on only one non-zero product term—this discrepancy seems to be an error.
Moreover, the provided algorithm (both in~\cite{llfs24} and its implementation code in~\cite{li24}) implicitly assumes that the accumulator retains an extra bit during significand alignment and no such variant for the alternative case is provided.
This assumption, however, does not hold for the NVIDIA V100 GPU, which does not employ any extra bits for significand alignment, in contrast to the A100 and T4 architectures~\cite{fasi2021}. Consequently, the algorithm in~\cite{llfs24} is not applicable to the V100 and similar GPUs, as explicitly stated therein.
It also gives an impression that FMA size is only linked to extra alignment bits whereas, in reality it also depends upon the extra bits allocated to accommodate carries, which we have shown below. 

The Algorithm~\ref{alg_ECB} also returns the FMA size along with the number of extra carry bits irrespective of whether extra alignment bits are present. 
Therefore, the proposed algorithm is applicable to GPUs with and without extra alignment bits, e.g. A100 and AMD employ extra alignment bits, and V100 do not. 
Moreover, it can also detect an FMA size as small as one. 
This is possible because when the Algorithm~\ref{alg_ECB} satisfiability condition is not true at $k=2$, the FMA size must be $k-1=1$. 
Otherwise, $k$ is incremented until the condition fails and the algorithm outputs $\NFMA=k-1$. 
The satisfiability condition holds as long as computation takes place within a single-block FMA and fails when normalisation and rounding become independent across different blocks. 
For example, if the FMA size is $4$ and $k=6$, then $c+\sum_{i=1}^{4}r_i$ is computed, normalised, and rounded within one block FMA, whereas $r_5+r_6$ is computed in the next block,
which returns a result $d$ not equal to $c+\sum_{i=1}^{6}r_i$. 
The proposed approach remains valid under any rounding mode applied after normalisation, which makes it reliable.
Note, we assume that the elements in each input vectors or matrices are distributed across multiple block FMAs in the same order as they are in the provided vectors or matrices. 
For instance, $\left[a_{q\NFMA+1},\dots,a_{(q+1)\NFMA}\right]$ and $\left[b_{q\NFMA+1},\dots,b_{(q+1)\NFMA}\right]$ are given to the $q$th block FMA, while $\left[a_{(q+1)\NFMA+1},\dots,a_{(q+2)\NFMA}\right]$ and $\left[b_{(q+1)\NFMA+1},\dots,b_{(q+2)\NFMA}\right]$ to the $(q+1)$th block FMA, and so on.
This assumption is reasonable in the context of matrix multiplication, as it enables simplified and efficient multiplication.
\begin{table*}[h]
\centering
\caption{Numerical Features of Matrix Multipliers
  in consumer-grade NVIDIA GPUs}
\label{table:results}
\begin{tabular}{c|c|cc|c|c|c|c|c|c}
\hline\hline
\multirow{2}{*}{Input format} & \multirow{2}{*}{Device} & \multicolumn{2}{c|}{Subnormal} & \multirow{2}{*}{$n_{\mathrm{eab}}$} & \multirow{2}{*}{$n_{\mathrm{ecb}}$} & \multirow{2}{*}{I.Norm} & \multirow{2}{*}{$\NFMA$} & \multirow{2}{*}{RM-BFMA} & \multirow{2}{*}{RM-MBFMA} \\ \cline{3-4}
                       &                         & \multicolumn{1}{c|}{In Support}  & Out Support &                       &                       &                           &                          &                           \\ \hline
\multirow{2}{*}{binary16 \cite{ieee19}}  & RTX 3060                & \multicolumn{1}{c|}{\ding{51}}   & \ding{51}   & 1                     & $\ge 3$                   & \ding{55} & 8                        & Truncate                       & Truncate                         \\ \cline{2-10} 
                       & Ada 1000                & \multicolumn{1}{c|}{\ding{51}}   & \ding{51}   & 1 & $\ge 3$                    & \ding{55} & 8                         & Truncate                        & Truncate                         \\ \hline
                       \multirow{2}{*}{bfloat16 \cite{inte18}}  & RTX-3060                & \multicolumn{1}{c|}{\ding{51}}   & \ding{51}   & 1                     & $\ge 3$                  & \ding{55}   & 8                        & Truncate                       & Truncate                         \\ \cline{2-10} 
                       & Ada 1000                & \multicolumn{1}{c|}{\ding{51}}   & \ding{51}   & 1 & $\ge 3$             &    \ding{55}    & 8                         & Truncate                        & Truncate                         \\ \hline

\multirow{2}{*}{TensorFloat32 \cite{nvid20}}  & RTX 3060                & \multicolumn{1}{c|}{\ding{51}}   & \ding{51}   & 1                     & $\ge 2$               &  \ding{55}    & 4                        & Truncate                       & Truncate                         \\ \cline{2-10} 
                       & Ada 1000                & \multicolumn{1}{c|}{\ding{51}}   & \ding{51}   & 1 & $\ge 2$              &   \ding{55}    & 4 & Truncate                        & Truncate                         \\ \hline\hline
\end{tabular}\\
\vspace{0.2em}
Note: The above results are for binary32 output. Binary16 output uses RN for both RM-BFMA and RM-MBFMA.\\
$n_{\mathrm{eab}}$ (Number of Extra Alignment Bits), $n_{\mathrm{ecb}}$ (Number of Extra Carry Bits) \& I.Norm (Immediate normalisation)
\end{table*}

\subsection{Rounding Mode in Compiling Results of Multiple Block FMAs (RM-MBFMA)}
Within a single block FMA operation, it is known from~\cite{hibr19, fasi2021} that truncation (or RZ) is typically employed when the output is in binary32, while RNE is used in binary16 and binary64 output modes.
However, the rounding behavior during the aggregation of results from two distinct block FMAs in a same sub-matrix or a tile requires further investigation.
Hence, we suggest setting $c=\pm(1+2^{-\pout+1})$ and $r_{N_{\mathrm{FMA}}+1}=\pm(2^{-\pout}+2^{-\pout-1}),~r_k=0~\forall k\neq {(\NFMA+1)}$. 
The output must be $d=\pm(1+2^{-\pout+1})$, $d=\{\pm(1+2^{-\pout+2})\}$, $d=\{(1+2^{-\pout+2}),-(1+2^{-\pout+1})\}$, and $d=\{(1+2^{-\pout+1}),-(1+2^{-\pout+2})\}$ for RZ/truncation, RNE, RU, and RD rounding modes, respectively.


\subsection{Normalisation Procedure \& Order of Addition between Two Block FMAs}
The warp matrix multiply accumulate (WMMA), a CUDA API to utilize tensor cores on NVIDIA GPUs, supports fixed tile sizes where the sizes depend upon the input data types and GPU architecture~\cite{nvid25a}. 
For matrices with dimension larger than these supported fixed size matrices, multiple WMMA operations are called to cover the entire input matrix. 
However, within the supported fixed-sized tiles, an inner product may not be computed via a single block FMA operation, instead multiple block FMA operations are performed to compute a single dot product. 
Therefore, if the shared dimension of the supported fixed sized tiles is denoted with $k_0$, there must be $k_0/\NFMA=k_{\mathrm{r}}\ge 1$, where $k_{\mathrm{r}}$ is an integer, block FMA operations to compute every inner product within a tile. 
Let us consider a case where the shared dimension of $A$, i.e., $k$, is equal to the shared dimension between tile multiplication i.e. $k=k_0$. Then we can write \eqref{eq_d1} as
\begin{align}
d=c+T_1+T_2+\dots,+T_{k_{\mathrm{r}}},
\end{align}
where $T_{\ell}=\sum_{i=(\ell-1)\NFMA+1}^{\ell\NFMA}r_i$.
We are interested in the order of addition operations between these terms. 
Since our experimental results, provided below, and also those reported in~\cite{llfs24} for NVIDIA GPUs, consistently yield $k_{\mathrm{r}}=2$, we focus on this case and propose a simple test. 
Assign $2^{j}, -2^{j}$ and $2^{-\pout-3+j}$ to  $c,~T_1$ and $T_2$, respectively, where $j\in\mathbb{N}_0$. 
We must have $d=2^{-\pout-3+j}$, $0$, and $0$ for $(c+T_1)+T_2,~c+(T_1+T_2)$ and $(c+T_2)+T_1$ ordering, respectively. 
Between the two cases yielding $d=0$, the input values can be permuted among \( c \), \( T_1 \), and \( T_2 \) to determine which ordering corresponds to each of the identical outcomes.

\section{Results}
Here we present the features available on RTX 3060 and Ada 1000 consumer-grade GPUs, as determined by our proposed generalised testing methodology.

\subsection{Subnormal Support}
Support for subnormal numbers both in input and output in above mentioned precision formats is available on both graphics cards. This complies with previous results of A100 from the Ampere family.
\subsection{Rounding Mode in a Block FMA (RM-BFMA)}
Binary32 output mode uses truncation instead of RNE, whereas RNE is the default rounding mode for binary16 output precision.

\subsection{Accumulator Features}
\subsubsection{Extra Alignment Bits}
In the binary32 output mode, results from both cards show that an extra bit is present when the significands are aligned during addition.
This indicates that the internal datapath for handling the significands of the addends is $25$ bits wide, same as what was reported by~\cite{fasi2021} for the A100.

\subsubsection{Extra Carry Bits} 
With an FMA size of $\NFMA$, akin to \eqref{eq_necb_k}, at most 
\begin{align}
\label{eq_ub}
n_{\mathrm{ecb.max}}=\floor{\mathrm{log_2}(\NFMA(2-2^{-\pin+1}))}
\end{align}
can be detected according to our definition of extra carry bits (see Sec.~II-B). 
It is also worth noting that while more extra carry bits may exist in matrix multiplier hardware units, detecting them would require a larger FMA size.
For both cards, Algorithm~\ref{alg_ECB}
detected $3$ extra carry bits in both graphics cards for binary16 and bfloat16 input, whereas for TensorFloat32 format as the input format, only $2$ extra carry bits are detected.
These results are consistent with the determined FMA size via \eqref{eq_ub}.  
The number of extra alignment and carry bits present suggests that the accumulator width, denoted in the format $\{n_{\mathrm{ecb}},24,n_{\mathrm{eab}}\}$, is at least $\{3,24,1\}$ for binary16 and bfloat16 as the input formats with the output as binary32, and $\{2,24,1\}$ for TensorFloat32 as the input format, with binary32 as the output format.

\subsubsection{Normalisation}
Running the test proposed for the case $\{eab, ecb\} = \{1, 1\}$ suggests that no immediate normalisation occurs within a single-block FMA operation on either the RTX 3060 or Ada1000 graphics card.

\subsection{FMA Size and Related Properties}

With \( e_{ab} = 1 \), the FMA size \( \NFMA \)—which depends only on the input precision—is found to be $8$ for binary16 and bfloat16, and $4$ for TensorFloat32, yielding \( k_{\mathrm{r}} = k_0 / \NFMA = 2 \) for all formats.  
This implies that two block FMA operations are used to compute each inner product between tiles from matrices \( {A} \) and \( {B} \).  
The test indicates that the addition order among \( c \), \( T_1 \), and \( T_2 \) follows \( (c + T_1) + T_2 \).  
When combining results from separate block FMAs, the rounding mode (RM-MBFMA) is truncation for binary32 output, and RNE for binary16 output.

\section{Conclusion}
We have proposed a numerical feature{} testing approach for the dedicated matrix-multiplier units available on many recent GPUs.
The proposed methodology can be applied to multiple input and output format combinations and is architecture independent. 
Unlike previous attempts, the proposed approach considers the dependencies among numerical features to lift this architecture dependency.
As a test case, we applied this methodology to characterize the numerical features of tensor cores on two consumer-grade NVIDIA GPUs—the RTX 3060 and the Ada 1000—using binary16, bfloat16, and TensorFloat32 input precision formats.  
The consumer-grade GPU based on the Ampere architecture shows identical features as that of the data centre GPU A100 except that RTX 3060 does not support binary64 in tensor cores.

In Part II, we plan to publish a comprehensive set of numerical features for matrix multipliers across GPUs from various vendors to facilitate cross-platform analysis and enhance portability. Moreover, this analysis will be extended to cover all available input precision formats, providing a user-friendly testing approach applicable to GPUs ranging from consumer-grade models to those featured in the TOP500 supercomputers. 
The code for this work is made} available on GitHub\footnote{\url{https://github.com/faiziktk/IEEE_HPEC2025_block_FMA_tests}} and it includes CUDA code for generating the results in Table~\ref{table:results} and MATLAB experiments with several simulated models of block FMA units.

\section{Acknowledgment}
We are grateful to Massimiliano Fasi for the insightful comments, which greatly improved the quality of this paper.
Both authors are funded by the EPSRC grant ``\emph{Informing Future Numerical Standards by Determining Features of Non-Standard Mathematical Hardware}'' with the project reference 151: \url{https://gtr.ukri.org/projects?ref=151}.

\bibliographystyle{IEEEtran}
\footnotesize{
\bibliography{IEEEabrv,references}

\begin{thebibliography}{10}
\providecommand{\url}[1]{#1}
\csname url@samestyle\endcsname
\providecommand{\newblock}{\relax}
\providecommand{\bibinfo}[2]{#2}
\providecommand{\BIBentrySTDinterwordspacing}{\spaceskip=0pt\relax}
\providecommand{\BIBentryALTinterwordstretchfactor}{4}
\providecommand{\BIBentryALTinterwordspacing}{\spaceskip=\fontdimen2\font plus
\BIBentryALTinterwordstretchfactor\fontdimen3\font minus
  \fontdimen4\font\relax}
\providecommand{\BIBforeignlanguage}[2]{{%
\expandafter\ifx\csname l@#1\endcsname\relax
\typeout{** WARNING: IEEEtran.bst: No hyphenation pattern has been}%
\typeout{** loaded for the language `#1'. Using the pattern for}%
\typeout{** the default language instead.}%
\else
\language=\csname l@#1\endcsname
\fi
#2}}
\providecommand{\BIBdecl}{\relax}
\BIBdecl

\bibitem{nvid25b}
\BIBentryALTinterwordspacing
NVIDIA, ``{NVIDIA} {Blackwell} architecture technical brief,'' 2025. [Online].
  Available: \url{https://resources.nvidia.com/en-us-blackwell-architecture}
\BIBentrySTDinterwordspacing

\bibitem{amd25}
\BIBentryALTinterwordspacing
AMD, ``Datasheet: {AMD} instrinct {MI355X GPU},'' 2025. [Online]. Available:
  \url{https://www.amd.com/content/dam/amd/en/documents/instinct-tech-docs/product-briefs/amd-instinct-mi355x-gpu-brochure.pdf}
\BIBentrySTDinterwordspacing

\bibitem{ieee19}
\emph{{IEEE} Standard for Floating-Point Arithmetic, {IEEE} {Std} 754-2019
  (revision of {IEEE} Std 754-2008)}.\hskip 1em plus 0.5em minus 0.4em\relax
  Piscataway, NJ, USA: Institute of Electrical and Electronics Engineers, Jul.
  2019.

\bibitem{ieee25}
I.~W. Group, ``Interim report on binary floating-point formats for machine
  learning,'' \url{https://github.com/P3109/Public/tree/main}, Jul. 2025,
  version 3.0.3.

\bibitem{hibr19}
B.~Hickmann and D.~Bradford, ``Experimental analysis of matrix multiplication
  functional units,'' in \emph{2019 IEEE 26th Symposium on Computer Arithmetic
  (ARITH)}, 2019, pp. 116--119.

\bibitem{fasi2021}
M.~Fasi, N.~J. Higham, M.~Mikaitis, and S.~Pranesh, ``Numerical behavior of
  {NVIDIA} tensor cores,'' \emph{PeerJ Computer Science}, vol.~7, p. e330,
  2021.

\bibitem{llfs24}
X.~Li, A.~Li, B.~Fang, K.~Swirydowicz, I.~Laguna, and G.~Gopalakrishnan,
  ``{FTTN}: Feature-targeted testing for numerical properties of {NVIDIA} \&
  {AMD} matrix accelerators,'' in \emph{2024 {IEEE} 24th International
  Symposium on Cluster, Cloud and Internet Computing ({CCGrid})}, 2024, pp.
  39--46.

\bibitem{vlpg25}
B.~Valpey, X.~Li, S.~Pai, and G.~Gopalakrishnan, ``An {SMT} formalization of
  mixed-precision matrix multiplication,'' in \emph{NASA Formal Methods}.\hskip
  1em plus 0.5em minus 0.4em\relax Cham: Springer Nature Switzerland, 2025, pp.
  360--379.

\bibitem{bhlt20}
P.~Blanchard, N.~J. Higham, F.~Lopez, T.~Mary, and S.~Pranesh, ``Mixed
  precision block fused multiply-add: Error analysis and application to {GPU}
  tensor cores,'' \emph{SIAM Journal on Scientific Computing}, vol.~42, no.~3,
  pp. C124--C141, 2020.

\bibitem{mika24}
M.~Mikaitis, ``Monotonicity of multi-term floating-point adders,'' \emph{IEEE
  Trans. Comput.}, vol.~73, no.~6, pp. 1531--1543, Feb. 2024.

\bibitem{li24}
\BIBentryALTinterwordspacing
X.~Li, ``Artifact for {FTTN},'' Feb. 2024. [Online]. Available:
  \url{https://doi.org/10.5281/zenodo.10673370}
\BIBentrySTDinterwordspacing

\bibitem{inte18}
{Intel Corporation}, ``{BFLOAT16}---hardware numerics definition,'' Available
  at
  \url{https://software.intel.com/en-us/download/bfloat16-hardware-numerics-definition}
  (accessed 15 July 2020), Nov. 2018, white paper. Document number
  338302-001US.

\bibitem{nvid20}
NVIDIA, ``{NVIDIA A100} tensor core {GPU} architecture,'' Available at
  \url{https://www.nvidia.com/content/dam/en-zz/Solutions/Data-Center/nvidia-ampere-architecture-whitepaper.pdf}
  (accessed 15 July 2020), pp. vi+76, 2020, {NVIDIA} whitepaper v1.0.

\bibitem{nvid25a}
\BIBentryALTinterwordspacing
------, ``{CUDA} {C}++ programming guide,'' 2025. [Online]. Available:
  \url{https://docs.nvidia.com/cuda/pdf/CUDA\_C\_Programming\_Guide.pdf}
\BIBentrySTDinterwordspacing

\end{thebibliography}
}

\end{document}